\documentclass[floatfix,twocolumn,nofootinbib,superscriptaddress,a4paper,preprintnumbers]{revtex4-1}

\usepackage{amsmath}
\usepackage{amssymb}
\usepackage{graphicx}
\usepackage[T1]{fontenc}
\usepackage{slashed}
\usepackage[utf8]{inputenc}
\usepackage{xspace}
\usepackage{color}
\usepackage{ulem}
\usepackage{array}

\newcommand{\AddrYantai}{Department of Physics, Yantai University, Yantai 264005, P. R. China}
\newcommand{\AddrPeking}{Center for High-Energy
Physics, Peking University, Beijing, 100871, P. R. China}
\newcommand{\AddrStaubb}{CAS Key Laboratory of Theoretical Physics, Institute of Theoretical Physics, Chinese Academy of Sciences, Beijing 100190, P. R. China}

\newcommand{\AddrNeimeng}{School of Physics and Electronic Information Inner Mongolia University for the Nationalities,Tongliao 028043,China}
\begin{document}


\title{Direct Detection of Axion-Like Particles in Bismuth-Based Topological Insulators}

\author{Tairan Liang}
\email{liangtr@ihep.ac.cn}
\affiliation{\AddrNeimeng}

\author{Bin Zhu}
\email{zhubin@mail.nankai.edu.cn}
\affiliation{\AddrYantai}


\author{Ran Ding}
\affiliation{\AddrPeking}

\author{Tianjun Li}
\affiliation{\AddrStaubb}

\begin{abstract}
In recent years a new field emerged in dark matter community and immediately attracted a multitude of theoriests and experimentalists, that of light dark matter direct detection in electronic systems. The phenomenon is similar with nuclear recoil in elastic scattering between dark matter and nucleus but with different kinematics. Due to the small energy gap, the electronic system can probe sub-GeV dark matter rather than nucleus target. In particular the absorption into materials can even detect ultra-light dark matter within mass around meV. In terms of the equivalence between optical conductivity and absorption cross section, axion detection can be computed in Bismuth-based topological insulators. It is found that topological insulator has strong sensitivity on axion and provides a complementary direct detection to superconductor and semiconductors. The novelty of topological insulator is that the thin film could even obtain the same sensitivity as superconductor.

\end{abstract}

\maketitle

\section{Introduction}
Axion~\cite{Peccei:1977hh,Weinberg:1977ma,Wilczek:1977pj} is an hypothetical particle introduced to solve the strong CP problem of QCD and it is also an interesting cold dark matter candidate with the relic density generated by misalignment mechanism~\cite{Preskill:1991kd}. Furthermore axions are known to have huge phase space density which is usually regarded as coherent field rather than a particle. The other interesting properties of axion are that it can form Bose-Einstein condensation~\cite{Sikivie:2009qn} in the cosmological evolution or a relaxation~\cite{Graham:2015cka} in order to solve the hierarchy problem. The detection of axion is really challenging because of its rather tiny couplings and masses that are inverted proportional to the PQ scale $f_a$. Compared with active search for WIMP such as LUX~\cite{daSilva:2017swg}, XENON~\cite{Aprile:2017aty} and PandaX~\cite{Cui:2017nnn}, the searches for axion in labs are fewer except ADMX~\cite{Simanovskaia:2016rcd}. Most of constraints for axion come from astro-physics where the stellar evolution in globular clusters, white dwarf cooling as well as HB stars set severe upper limits on the couplings and masses. In other words $f_a$ has a lower bound $f_a>3\times 10^9$ GeV.

For WIMP the traditional direct detection strategy~\cite{Gelmini:2015zpa} is looking for nuclear recoil energy from the elastic or inelastic scattering process between WIMP and nucleus. For a typical WIMP with mass around $50$ GeV, the corresponding recoil energy is $100$ keV which is compatible with the threshold of certain materials i.e. Xe, Ge in direct detection experiment so that it can capture the incoming dark matter kinematic energy. It in turn generally requires the WIMP mass should be larger than $1$ GeV.

When dark matter is lighter than $1$ GeV i.e. axion which is a theoretically motivated but largely unexplored paradigm. The kinematics of scattering into target undergo an interesting transition where the electron recoil energy becomes more significant than that of nucleus target. As a consequence electronic system becomes more attractive for searching for light dark matter. The existing experimental proposals in electronic system contain the ionization of noble liquid~\cite{Essig:2011nj,Essig:2012yx}, single electron event in semi-conductor detector~\cite{Essig:2015cda,Essig:2017kqs}, dark matter scattering in scintillating targets~\cite{Derenzo:2016fse}, graphene~\cite{Hochberg:2016ntt}, superconductor~\cite{Hochberg:2015pha,Hochberg:2015fth}, superfluid helium~\cite{Schutz:2016tid,Knapen:2016cue},	
magnetic bubble chambers~\cite{Bunting:2017net} and three-dimensional Dirac Materials~\cite{Hochberg:2017wce}. In particular the absorption in superconductor~\cite{Hochberg:2016ajh}, semiconductor~\cite{Hochberg:2016sqx,Bloch:2016sjj} as well as molecules~\cite{Arvanitaki:2017nhi} provides an alternative strategy for detecting ultra-light dark matter which was proposed in~\cite{Pospelov:2008jk,Derevianko:2010kz} .

In this paper we focus on topological insulator~\cite{Hasan:2010bin,Qi:2011zya,Fu:2007uya} to host the axion absorption which attracts lots of attentions in condensed matter community. For much of the twentieth century, classification of phases of matter are understood by the principle of spontaneous symmetry breaking principle, called Landau approach. However with the discovery of integer and fractional quantum Hall effect (QHE) in 1980s, a new order of state appeared which depends on topological quantum numbers rather than symmetry transformation. In the last few years it has been even found that such a topological order also occurs in specific three-dimensional topological insulator. This type of material is what we use throughout this paper. In spite of QHE, the role of external magnetic field is now replaced by intrinsic strong spin-orbit interaction. As a result, topological insulators are host to topologically protected metallic surface states with bulk remaining insulator. The physical content of surface state in topological insulator is identified as chiral Dirac fermion through its dispersion relations which is the main property of topological insulator and make themselves robust against non-magnetic impurities. The distinction of bulk and surface states plays a crucial role in axion detections. The reason why topological insulator attracts lots of attention is that it provides not only an intellectual appeal of topological order without external magnetic field,  but a promise candidate for spintronics and quantum computing~\cite{Kitaev:2005dm,Vijay:2015zia}. For now its application can be extended to detect axion-like dark matter or bosonic super-WIMPs. In terms of the excellent equivalence between optical conductivity and absorption cross section, the project sensitivity can be computed analytically.   We want to study in the following for the first time explicitly the impact of topological insulator on the axion absorption. As example we consider $\text{Bi}_2 \text{Se}_3$ and its alloys that are called second generation topological insulators for historical reasons~\cite{Hsieh:2009fak,Neupane:2014dfa}. Since Bismuth-based topological insulator has a meV energy gap, it can even probe further smaller axions compared with superconductor and semi-conductor.

The rest of paper is organized as follows: in section II
we briefly review the particle physics content of axions and layout the relevant vertex for our search. In section
III we show the analytical derivation of equivalence between
optical conductivity and absorption cross section. In section IV we
present the optical conductivity of topological crystal insulator, topological
insulator thin film respectively. Afterwards we perform a numerical study
of projected sensitivity. We
conclude in section V.

\section{Partial Visit to Axion}
Axion is originally introduced to solve the strong CP problem. The standard QCD lagrangian contains a additional term that forsees CP violation.
\begin{align}
\mathcal{L}=-\frac{g_s}{32\pi^2}\bar\theta G_{\mu\nu}\tilde G^{\mu\nu}
\end{align}

where $G_{\mu\nu}$ is gluon and $\tilde G^{\mu\nu}$ is dual to gluon field. The QCD topological effect as well as electroweak contribution give rise to $\bar\theta$.

\begin{align}
\bar\theta=\theta-\arg\det M
\end{align}
where $M$ is quark mass matrix and $\theta$ is topological term that characterizes degenerate QCD ground vacuum. Though $\theta$ is not specified by standard model itself, people usually expect it to be around 1 by naturalness. As we all know the topological theta term $\theta$ plays no role in the perturabtive calculation. However CP-violating effect in $\bar\theta$  generates non-vanishing neutron electic dipole moment (EDM).  The current limits from EDM experiments show that $\bar\theta$ is smaller than $10^{-10}$. Such a fine-tuned parameter motivates Peccei and Quinn~\cite{Peccei:1977hh} to propose a enhanced $U(1)$ symmetry to protect it. The Peccei-Quinn proposal is then realized as existence of a pseudo-scalar i.e. axion by Weinberg~\cite{Weinberg:1977ma} and Wilczek~\cite{Wilczek:1977pj} with axion being a Goldstone boson of spontaneously broken $U(1)$ symmetry. In Weinberg-Wilczek model, $\bar\theta$ is replaced by a dynamical pseudo-scalar field
\begin{align}
\bar\theta=\frac{a(x)}{f_a}
\end{align}

where $f_a$ is the scale that $U(1)$ symmetry is spontaneously broken. Thus the mass and couplings of axion are inversely proportional to $f_a$.  This type of axion is called QCD axion. The best accepted scenario is that $f_a$ is close to electroweak scale such that axion could be detected at collider. Neverthless such a choice is quickly excluded by experiments. Therefore we are left with insisible axion models  in which $f_a$ is far larger than electroweak scale called invisible axion models.

The invisible axion is a natural candidate of dark matter particle. The corresponding relic density of ultralight axion is set up by misalignment mechanism~\cite{Preskill:1991kd}.  Compared with relatively heavy WIMP dark matter, the typical property of axion-like particle dark matter is that it has a much higher phase space density. Therefore the search for axion is no longer a single hard particle scattering but a coherent effects of entire field which makes axion absorption in material plausible.

Depending on different mass, coupling relation, there are two benchmark axion models: KSVZ~\cite{Kim:1979if,Shifman:1979if} and DFSZ~\cite{Zhitnitsky:1980tq,Dine:1981rt}. Several interactions in these benchmark models are allowed and satisfied up to some certain relation: axion-quark, axion-photon, axion-electron. Their coulings are highly model dependent and provide different search strategies for axion. For example the axion-photon coupling gives rise to primakoff effect which can lead to the conversion of an axion into a photon in the presence of a large scale magnetic field. This method was first proposed by Sikivie~\cite{Sikivie:1983ip} and called cavity method which can even explore more complicated axion models than DFSZ and KSVZ.

\begin{align}
\mathcal{L}=-g_{a\gamma\gamma}a F_{\mu\nu}\tilde F^{\mu\nu}=
=-g_{a\gamma\gamma}a E\cdot B
\end{align}
Here $F$ is the electromagnetic field stength tensor, $\tilde F$ is its dual. The coupling constant $g_{a\gamma\gamma}$ is given
\begin{align}
g_{a\gamma\gamma}=\frac{\alpha}{2\pi f_a}\left(\frac{E}{N}-\frac{2}{3}\frac{4+z}{1+z}\right)
\end{align}
where $E$ and $N$ are the electromagnetic and color anomaly of the axial current associated with the axion field respectively. In addition $z=m_u/m_d$ is the quark mass ratio. In the context of DFSZ model, $E/N=8/3$ so that $g_{a\gamma\gamma}=0.36$; while in KSVZ model, $E/N=8/0$ so that $g_{a\gamma\gamma}=-0.97$. For generic $E/N$, we obtain the axion-like model that we concern in this paper.  The large scale magnetic fields can be obtained either from laboratory or astro-physics. Both of them put severe constraints over the mass and coupling of axion models which will be analyzed in the following section.  In our paper we use a different promising approach to search axion, i.e. axion-electron interaction,

\begin{align}
\mathcal{L}=-g_{aee}\partial_{\mu} a \bar e\gamma^{\mu}\gamma_5 e
\label{eqn:axoele}
\end{align}

The interaction in equation~(\ref{eqn:axoele}) is a derivative interaction so that it preserves the shift symmetry of axion $a\rightarrow a+a_0$. In DFSZ, $g_{aee}=4.07\times 10^{-11} $; while in KSVZ, $g_{aee}$=0. The yukawa type coupling $g_{aee}$ determines the strength between axion and electrons. In contrast with conventional direct detection with nuclear recoil target, the electron system has a relatively low threshold and is well explored in condensed matter physics. As a result the ultra-light dark matter can scatter off electrons in materials easily and provides a clean signal. In particular when the axion is as light as meV, the electron system can absorb the incoming axion , then evolves from valence band to conduction band. The transition can be regarded as a smoking gun for axion detection.
The relevant content from particle physic to characterize axion model is thus the mass $m_a$ and coupling $g_{aee}$ to electron, both of which will be constrained within topological insulators.

Finally we mention that even though we assume there is only one interaction at tree level i.e. axion-electron is allowed, it will automatically generate a loop-induced axion-photon interaction,
\begin{align}
\frac{\alpha}{8\pi}\frac{g_{aee}}{m_e} a F_{\mu\nu}\tilde F^{\mu\nu}
\label{eqn:loop}
\end{align}

The loop-induced interaction in equation~(\ref{eqn:loop}) brings the constraints for axion-photon into our model which must be considered seriously.
\section{Dark Matter Absorption Formulation}
The purpose of this section is to demonstrate that superconductors, semiconductors as well as topological insulators can be used to probe ultralight axion via absorbing process. We now turn to computing the rate of DM absorption in a material. The total axion absorption rate R per unit mass pert unit time is

\begin{align}
R=\frac{1}{\rho}\frac{\rho_{a}}{m_{a}} \langle n_e \sigma_{\text{abs}} v_{\text{rel}}\rangle
\end{align}
where $\rho$ is the mass density of the target material, $n_e$ is the number density of target systems, $\sigma_{\text{abs}}$ is the axion absorption cross section on electrons, $v_{\text{rel}}$ is relative velocity between axion and electrons and $\rho_{a}$ is local dark matter mass density which is set to be $0.3 ~\text{GeV}/\text{cm}^3$.

Given target number and dark matter mass, the computable quantity is thus the velocity averaged cross section.  In Condensed Matter Physics (CMP), one can treat the target material i.e. topological insulator as free electrons with certain Fermi energy $E_{F}$. The corresponding cross section involving phonon emisson is written as follows,

\begin{align}
\langle n_e \sigma_{\text{abs}} v_{\text{rel}}\rangle
=\int \frac{d^3 Q}{(2\pi)^3} \frac{\langle\mathcal{M}^2\rangle}{16E_1 E_2 E_3 E_4} S(q, Q)
\end{align}

One can calculate the matrix element $\mathcal{M}$ in terms of Feynman diagram method by treating phonon as a scalar field $\Phi$ with the dimensionless yukawa coupling
\begin{align}
y_{\Phi}=\frac{C_{\Phi}Q}{\sqrt{\rho}}
\label{eqn:yphi}
\end{align}

The equation~(\ref{eqn:yphi}) shows that the cross section is computable once we fix the value of coupling $C_{\Phi}$.  However it is totally different from particle physics that $C_{\Phi}$ can not be determined by underlying theory but obtained by matching onto data. The constant $C_{\Phi}$ which connect  photon absorption with dark matter absorption plays a central role in determining the rate. We in this section manage to prove that there is actually no need to acutally solve for $C_{\Phi}$ but write the dark matter cross section in terms of photon absorption cross section. The basic procedure in~\cite{Hochberg:2016ajh,Hochberg:2016sqx,Bloch:2016sjj,Hochberg:2017wce}  is illustrated as follows:

\begin{itemize}
\item We first compute the matrix elements of $\mathcal{M}_{\gamma}$ and
$\mathcal{M}_{\chi}$ in terms of feynman diagrams respectively
\begin{align}
\mathcal{M}_{\gamma}^2&\sim \frac{4e^2}{3}\frac{C_{\Phi}^2}{\rho}\frac{Q^4}{\omega^2}=g[Q,\omega] C_{\Phi}^2
\end{align}
where $g[Q,\omega]$ is the characterisic constant for photon absorption. Correspondingly we also have $\mathcal{M}_{\chi}^2=f[Q,\omega]C_{\Phi}^2$.

\item Matching the two matrix elements, $C_{\Phi}$ is cancelled automatically,
\begin{align}
&\mathcal{M}_{\chi}^2=\frac{f[Q,\omega]}{g[Q,\omega]}\mathcal{M}_{\gamma}^2
  =\kappa \mathcal{M}_{\gamma}^2\nonumber\\
&\langle n_e \sigma_{\text{abs}} v_{\text{rel}}\rangle_{\chi}=\kappa^2
\langle n_e \sigma_{\text{abs}} v_{\text{rel}}\rangle_{\gamma}
\end{align}
where $\kappa^2$ is the ratio between factors without $C_{\Phi}$ which is model-dependent parameter.  For axion it can be obtained from the equation~(\ref{eqn:axoele}),
\begin{align}
\kappa^2=\frac{3m_a^2}{4m_e^2}\frac{g_{aee}^2}{e^2}
\end{align}
\end{itemize}

Thus we can relate them with each other simply. Furthermore we can relate the dark matter absorption cross section with optical conductivity. As we know the current-current correlation function characterizes the complex optical conductivity
\begin{align}
\Delta =-i\sigma
\label{eqn:delta}
\end{align}

We mention that the actual current-current correlation function is $\Delta_{\mu\nu}$ which can be reduced to $\Delta$ in a isotropic medium. In terms of optical theorem, the current-current correlation function can be cut into absorption process rate
\begin{align}
\langle n_e\sigma_{\text{abs}} v_{\text{rel}}\rangle_{\gamma}
=-\frac{\text{Im}\Delta}{\omega}
\label{eqn:opt}
\end{align}

From equation~(\ref{eqn:delta}) and (\ref{eqn:opt}) we have
\begin{align}
\langle n_e\sigma_{\text{abs}} v_{\text{rel}}\rangle_{\gamma}
=-\frac{1}{\omega}\text{Im}(-i\sigma\omega)
=\text{Im}(i(\sigma_1+i\sigma_2))=\sigma_1
\label{eqn:cross}
\end{align}
The equation~(\ref{eqn:cross}) is the bridge connecting particle physics with condensed matter physics. In terms of it, there is no need to compute the actual absorption cross section. The experimental confirmation of optical conductivity is good enough to capture all the relevant information of axion detection. For a given material such as topological insulator, all that we should do is to compute or refer to its optical conductivity. The dark matter rate $R$ is now rewritten as follows

\begin{align}
R=\frac{1}{\rho}\frac{\rho_{\chi}}{m_{\chi}}\kappa^2 \sigma_{1}
\end{align}

\section{Optical Conductivity of Topological Insulators and Projected Sensitivity Reach}

The strong $3D$ topological insulator $\text{Bi}_2 \text{Se}_3$ and its alloys share similar crystal structure which consists of alternating hexagonal monatomic crystal planes stacking in ABC order. One of the most interesting aspects of Bismuth based topological insulators is their unique electronic structure. The materials exhibit semiconducting behavior with bulk band gaps approximately $300$ meV. Such a low threshold allows topological insulator to probe even tinier mass axions.

\begin{figure} [tb]
\begin{center}
\includegraphics[width=0.50\textwidth]{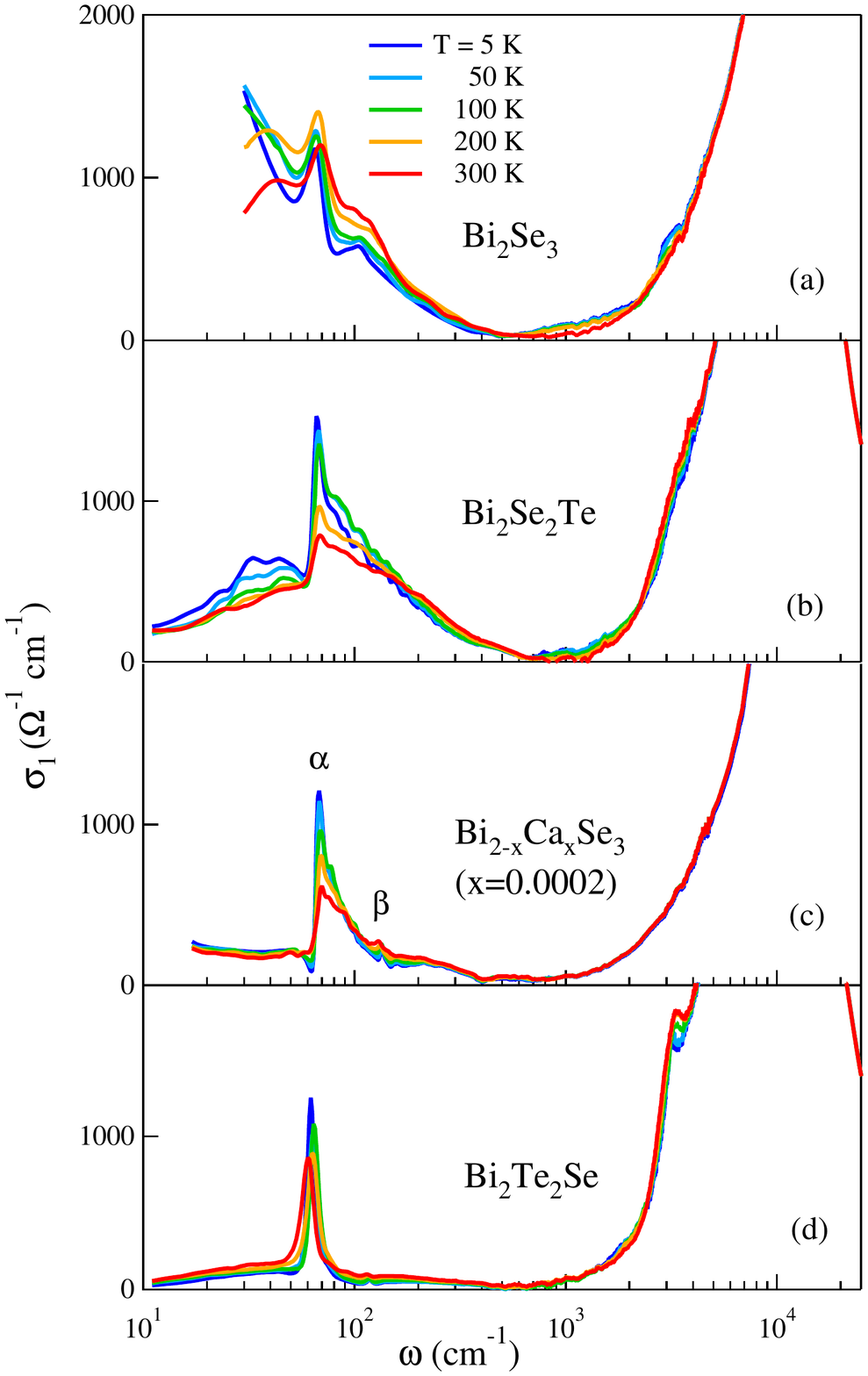}
\end{center}
\caption{The optical conductivity of $\text{Bi}_2\text{Se}_3$}, $\text{Bi}_2\text{Se}_2\text{Te}$ (b), $\text{Bi}_{1.9998}\text{Ca}_{0.0002}\text{Se}_3$ (c), $\text{Bi}_2\text{Te}_2\text{Se}$ (d)from $10$ to $24000\text{cm}^{-1}$ in Ref. The peak in the figure corresponds to phonon emission thus called intra-band. The $\alpha$ and $\beta$ infrared-active phonon modes are
indicated in panel (c). transition.
\label{fig:optical}
\end{figure}

Since the electronic transport experiment can not distinguish surface state from bulk contributions, the early experimental confirmation of topological insulator are performed in terms of Angle Resolved Photeoemission (ARPES) experiment. It demonstrated that $\text{Bi}_2\text{Se}_3$ and $\text{Bi}_2\text{Te}_3$ are strong three dimensional topological insulator with a single Dirac cone.  The optical conductivity of them are measured through the combination of reflectivity and Kramers-Kronig transformation~\cite{2012PhRvB..86d5439D}. In figure~\ref{fig:optical}, we show the measurements of the optical conductivity in topological crystal insulators including $\text{Bi}_2\text{Se}_3$, $\text{Ca}$ doped $\text{Bi}_2\text{Se}_3$($\text{Bi}_{2-x}\text{Ca}_x\text{Se}_3$) and alloys $\text{Bi}_2\text{Te}_2\text{Se}$ and $\text{Bi}_2\text{Se}_2\text{Te}$.

\begin{figure} [tb]
\begin{center}
\includegraphics[width=0.50\textwidth]{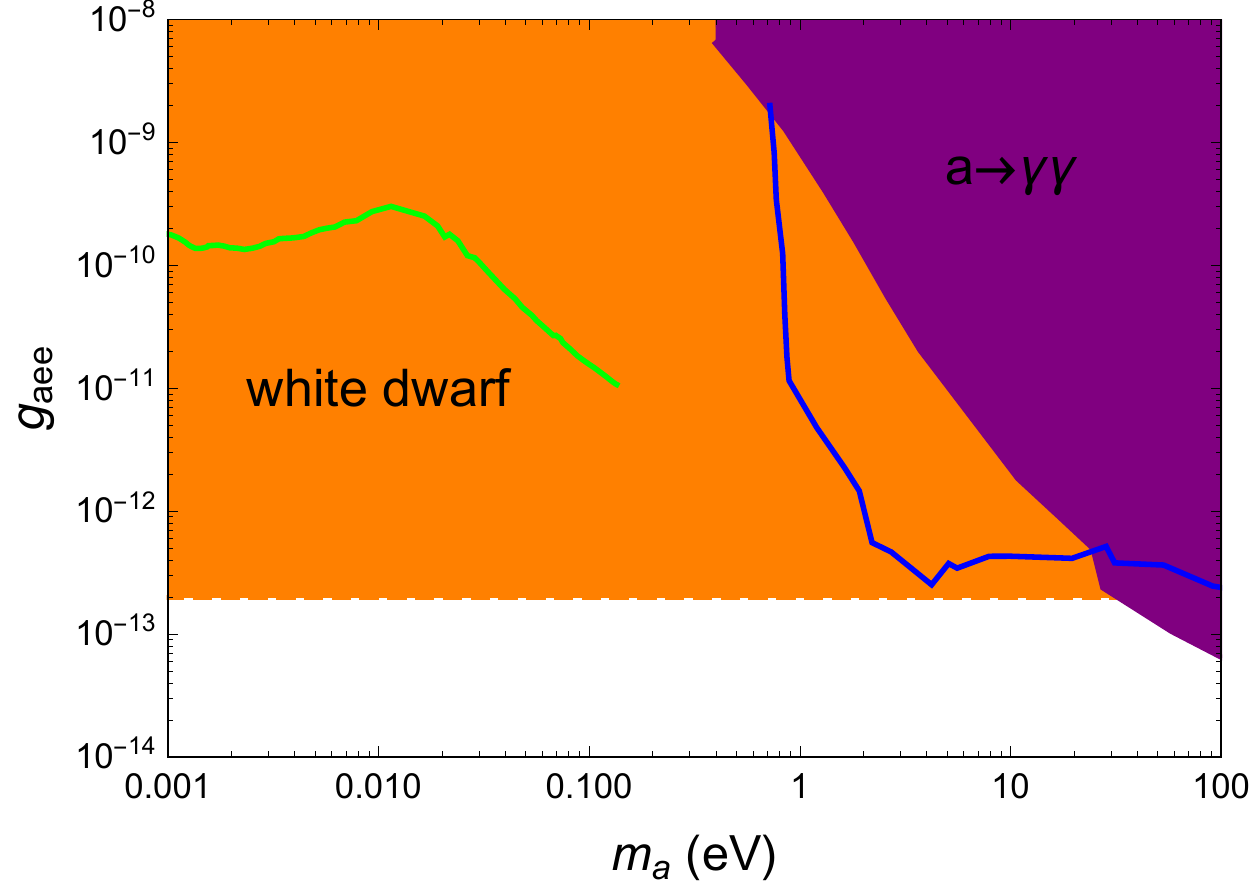}
\end{center}
\caption{The projected sensitivity is given explicitly with green line. In comparison with semi-conductor, we show the projected sensitivity of Ge with blue line. The shaded regions correspond to various constraints from astrophysics and particle physics.}
\label{fig:sen}
\end{figure}

The interband transition corresponds to the small bump which occurs at $3000~\text{cm}^{-1}$. It is too difficult to identify because of its superposition with the huge triplet electronic excitation present above $10000~\text{cm}^{-1}$. The energy scale of interband transition is overlap with semiconductor. Thus it is not our focus in this paper. The sub-THz region in~\cite{2012PhRvB..86d5439D} has two peaks corresponding to phonon emission. In terms of optical conductivity, we can obtain the projected sensitivity for $1$ kg $Bi_2 Se_3$ $1$ year. In figure~\ref{fig:sen},  we show the projected sensitivity explicitly. The green line corresponds to the $Bi_2 Se_3$  sensitivity while the red one is obtained through Ge absorption. It is easy to find that the topological insulator provides a complementary detection to semi-conductor where meV axion can be detected.

The shaded region contains various constraints from stellar or particle physics. Here we only present the two strongest constraints:
\begin{itemize}
\item The interaction between pseudo-scalar axion and electrons introduces compton-like scattering and eletromagnetic bremsstrahlung which in turns leads to constraints from star observation. In terms of I-band brightness of tip of red-giant brach, the limit on axion-electron yukawa $g_{aee}$ could be obtained~\cite{Viaux:2013lha}

\begin{align}
g_{aee}&<2.6\times 10^{-13} \; (68\%\, \text{CL})\nonumber\\
g_{aee}&<4.3\times 10^{-13} \; (95\%\, \text{CL})
\end{align}
In this paper we choose a moderate value i.e. $g_{aee}<3\times 10^{-13}$. This is the most important constraint than any other constraint such as Xenon.

\item As is mentioned in equation~(\ref{eqn:loop}), the loop induced coupling from $g_{aee}$ leads to dangerous $a\rightarrow\gamma\gamma$ on the extragalactic background light, early reionization, and X-rays~\cite{Arias:2012az}.
\end{itemize}

It is easy to find that topological insulator can detect meV scale axions due to its small mass gap which provides a complementary search strategy for semi-conductor. The low-sensitivity comes from the underlying similarity between semi-conductor and topological insulator that both of them have fewer electrons than superconductor. Thus it should take more topological insulator to obtain the same sensitivity as superconductor.

The intrinsic problem of topological crystal insulator is that it is really hard to distinguish its surface optical conductivity from extrinsic conductivity. This is mainly because the surface effect suffers from a non perfect stoichiometry due to Se vacanicies and Te defects. That is to say, the behavior of optical conductivity in topological crystal insulator are almost similar with that of semiconductor. That explains the similarity of topological insulator and semi-conductor. The only difference appears on the numerical discrepancies on projected sensitivity and mass range. That motivates us further explore the optical properties of topological insulator by using thin films. It has been proven that the conductance of topological insulator thin films are nearly independent of free carrier contribution on film thickness in The Valdes Aguilar et al in PRL . The useful quantity to characterize thin film in experiment is conductance $G$,

\begin{align}
G=\sigma d
\end{align}

where $d$ is the thickness of the thin film. From figure $2$, it is easy to find the surface states conductance does not vary along with the thickness, while the $3$-dimensional property scales with the thickness. This reveals that the free-carries contribution comes mostly from surface states. We finally obtain a framework where axion can be absorbed by surface chiral fermions. The topological insulator thin film manifests its optical conductivity with Drude term in~\cite{2013arXiv1302.4145S}, where the author caculate the optical conductivity for the surface excitations of a topological insulator as a function of the chemical potential $\mu$. When the absorbed photon energy is larger than $2\mu$, the optical conductivity of surface state is universal and not dependent on the structure,

\begin{align}
\sigma=\frac{e^2\pi}{8h}
\end{align}

which is four times smaller than universal conductance in mono-lyer graphene. When the energy is smaller than $2\mu$, we obtain a Drude peak of topological insulator surface state,

\begin{align}
\sigma=\frac{e^2}{2h}\frac{k_F l_{el}}{(\omega\tau)^2+1}
\end{align}

where $l_{el}$ is the elastic mean free path with $l=v\tau$. We also easily obtain the figure for optical conductivity with incoming photon energy. Thus its signal sensitivity is similar with superconductor or mono-layer graphene. As a result the topological thin film could obtain the same sensitivity as superconductor. This is the biggest achievement that topological insulator that its crystal form can detect meV dark matter while thin film form can obtain better sensitivity.
Besides that, due to its two-dimensional behavior, topological thin film could obtain the directional information for the emitting phonon. That can help us remove the background. We leave this topic in the forth-coming project.

\section{Conclusion}

In this paper, we propose a novel approach to detect ultra-light axions. Based on the interaction between electrons and axions, topological insulator becomes a well material to absorb axions. Its low mass gap allows us to detect meV axions which provide a complementary channel to semi-conductors. Another advantage of topological insulator is that its thin film form has Drude-like optical conductivity thus allows us to obtain better sensitivity than semi-conductors. That is to say, within topological insulator, we have two different searches for axions.

\section*{Acknowledgements}
We thank Tongyan Lin for interesting discussions and technical support for the background estimations. Bin especially thanks Qiaoli Yang for clarifying various issues of axions to me. Tairan Liang is supported by the National Natural Science Foundation of China (No.11705098).
Bin is supported by the National Science Foundation of China (11747026 and 11575151).

\bibliography{lit}

\end{document}